\documentstyle[12pt,aaspp4]{article}

\begin{document}

\def\ltsima{$\; \buildrel < \over \sim \;$}
\def\lsim{\lower.5ex\hbox{\ltsima}}
\def\gtsima{$\; \buildrel > \over \sim \;$}
\def\gsim{\lower.5ex\hbox{\gtsima}}
\def\sinc{\hbox{sinc}}

\title{On the relationship between the periodic and aperiodic variability of
accreting X--ray pulsars}

\author{Davide Lazzati}
\affil{Osservatorio Astronomico di Brera \\
Via E. Bianchi, 46 --
22055 Merate (Lecco), Italy \\
and \\
Universit\`a degli Studi di Milano - sede di Como \\
Via Lucini, 3 --
22100 Como, Italy \\
{\it E--mail}: lazzati@merate.mi.astro.it}

\author{Luigi Stella$^1$}
\affil{Osservatorio Astronomico di Roma \\
Via dell'Osservatorio, 2 --
00040 Monteporzio Catone (Roma), Italy \\
{\it E--mail}: stella@coma.mporzio.astro.it \\
$^1$ Affiliated to the International Center for Relativistic Astrophysics}

\begin{abstract}

Besides the narrow peaks originating from the periodic signal, 
the power spectra of accreting X-ray pulsars display continuum 
components usually increasing towards low frequencies; these   
arise from the source aperiodic variability. Most studies 
so far adopted the view that the periodic and aperiodic variations
are independent. However any aperiodic variability in the emission 
from the accretion column(s) towards the magnetic neutron star 
should be modulated at the X-ray pulsar period, by 
virtue of the same rotation-induced geometric effects which give
rise to the periodic signal. We develop here a simple shot noise model 
to test the presence of a coupling between the periodic and aperiodic
variability of X-ray pulsars. The model power spectrum is fit to the 
power spectra of three X-ray pulsars, Vela~X-1, 4U~1145-62 
and Cen~X-3 observed with EXOSAT. In the first two cases we find that a highly
significant coupling is present, as testified by a substantial
broadening in the wings of the power spectrum peaks due to the 
periodic modulation. We find also that these wings can mimick 
the presence of a knee in the continuum power spectrum components
around the fundamental of the periodic modulation, therefore 
questioning the correlation reported by Takeshima (1992) between 
the X-ray pulsar frequency and the knee frequency, beyond which the 
continuum power spectral component steepens.

\end{abstract}

\keywords{X-rays: binaries --- Pulsars --- Stars: accretion --- Stars: neutron
--- Stars: X-rays }

\section{Introduction}
\label{intro}

That fast aperiodic X-ray flux variations are present in accreting magnetic 
neutron stars became clear with the discovery of  4.4~s pulsations from
V0332+53, a transient X-ray binary that had  tentatively been classified as
a black hole candidate based on  the similarity of its short term
variability to that of Cyg~X-1 (Tanaka et al. 1983, Stella et al. 1985, 
Makishima et al. 1990).  
A number of studies of the aperiodic variations of X-ray pulsars has
been  carried out since. In most cases these studies have relied upon  power
spectrum techniques in a way that parallels the application to 
non-pulsating X-ray binaries. Continuum power spectrum  components arising
from the source  aperiodic variability are often identified, fit to
analytic models  and used to characterise the short term variations in
relation to other  source properties (such as spectral or luminosity state). 
These continuum power spectrum components provide also an important element
for  classifications across different types of accreting X-ray sources (see
e.g. van der Klis 1995 and references therein). 

In addition to the narrow peaks originating from the periodic modulation, 
the power spectra of many X-ray pulsars show distinct noise  component(s)
increasing towards low frequencies. In several of the slower  wind-fed
X-ray pulsars the power increases in a power law-like fashion down  to the
lowest frequencies sampled (usually $10^{-4}-10^{-3}$~Hz). Some  of the
faster disk-fed X-ray pulsars show instead a (nearly) 
flat-topped noise component at low frequencies which steepens 
above a  knee at a
frequency of $\nu_{knee}$ (Nagase 1989; Soong and Swank 1989). 
To date the most detailed and comprehensive study of the
power spectra of X-ray pulsars has been carried out by Takeshima (1992), 
based on Ginga observations. This author reports that a
flat-topped or a moderately steep  noise component is present in  virtually 
all X-ray pulsars shortwards of the peak corresponding to the fundamental 
of the periodic modulation.  In many cases the knee above which the power 
spectrum steepens lies close to the base of this peak, such that a strong 
correlation is found between the pulse frequency $\nu_p$ and the knee 
frequency $\nu_{knee}$,  centered around the relation $\nu_{s} \simeq
\nu_{knee}$  (Takeshima 1992). 
 
Several other continuum power spectrum features extending over a  more
limited range of frequencies (usually a decade or less) are 
observed in X-ray pulsars: these include broad bumps,
wiggles and steep very low frequency excesses.  Broad power spectrum
peaks testifying to the presence of  quasi periodic oscillations have
been clearly revealed in some cases, notably Cen~X-3
(Nagase 1989; Soong and Swank 1989), EXO~2030+375 (Angelini 
et al. 1989), 4U~1626-67 (Shinoda et al. 1990) and A0535+262 
(Finger et al. 1996). The
continuum power spectrum components of individual X-ray  pulsars have
been observed to change in relation to a variety of phenomena such as 
source flux, orbital phase and spin-up versus spin-down 
(Angelini 1989; Angelini et al. 1989; Belloni \& Hasinger 1990a; Parmar
et al. 1989). 

Most studies so far implicitly adopted the view that 
the periodic and aperiodic variability
of X-ray pulsars are independent, such that the light curves 
are given by the sum of the two corresponding signals; similarly
the power spectra are given by the sum of the individual power spectra of 
the  two signals (see e.g. Angelini et al. 1989). 
On the other hand it is well known that X-ray pulsars
generate their periodic signal by virtue of geometric effects arising
from the rotation-induced motion of
the accretion column(s) through which the inflowing material is 
funneled onto the magnetic poles of the neutron star. Any aperiodic 
variation in the emissivity of the accretion column(s) should therefore 
be modulated by the periodic signal as well. Accordingly, Makishima (1988) 
suggested an approach in which the aperiodic variability of an X-ray 
pulsar is multiplied by (as opposed to summed to) the periodic modulation.
The coupling between the aperiodic and periodic variations, if present, 
is expected to alter the shape of the power spectrum, due to the 
convolution of the corresponding Fourier transforms (see also 
Burderi et al. 1993).

In this work, we develop a simple model for testing the  presence of a
coupling between the  aperiodic variability of X-ray pulsars and their
periodic signal.  We adopt a suitable shot noise model to work out
analytically the  expected power spectrum for an arbitrary coupling factor.
The model power spectrum obtained in this way is then fit to the power
spectra  of several X-ray pulsars. In two out of three cases we find that a 
significant coupling is present between the shots and the periodic signal,
as testified by a substantial broadening in the wings of the power spectrum
peaks arising from the periodic modulation. In these two cases the 
continuum power still increases shortwards of the periodic modulation 
peak(s) and the broad wings mimick the presence 
of a knee in the power continuum close to $\nu_p$. 
By analogy this suggests that also in other X-ray 
pulsars with extended red noise components the power spectrum knee 
close to $\sim \nu_p$ might be an artifact produced by the broad wings of the 
coherent modulation peak(s); therefore, the reported correlation 
between $\nu_p$ and $\nu_{knee}$ appears to be questionable. 

Our paper is organised as follows: Section~\ref{modello} describes 
the analytic shot noise model that we have developed; the application 
of this model to the power spectra from the EXOSAT light curves
of selected X-ray pulsars is described in Section~\ref{applicazione}; 
our results are discussed in Section~\ref{discussione}. 

\section{The shot noise model}
\label{modello}

Shot noise models provide a useful mathematical description of  
the accretion flow inhomogeneities that 
are believed to generate the variability of 
accreting compact stars (Terrell 1972; Weisskopf et al. 1975; 
Sutherland et al. 1978; Shibazaki \& Lamb 1987; Elsner et al. 1987;
Abramowicz et al. 1991).
Individual shots are often supposed to be representative of 
the emission from self-luminous blobs or clumps in the accretion process. 
In the context of accreting X-ray pulsars, models involving 
density fluctuations in the accreting plasma have 
been discussed by Burderi et al. (1993) and Hoshino \& Takeshima (1993). 
 
We assume that the X-ray pulsar signal consists of two different 
components: (1) a deterministic periodic component:

\begin{displaymath}
f_1(t) = A + \sum_{n=1}^N B_n \sin{(n\omega_0 t + \phi_n)} \qquad ,
\end{displaymath}

\noindent
where $\omega_0\ =\ 2\pi / P$ is the fundamental of the pulsed signal 
of period $P$, $N$ the number of its harmonics and $\phi_n$ the
corresponding phases; (2) an aperiodic component consisting of the
superposition of  a number of similar shots, which gives rise to the red
noise power spectra observed in most X-ray pulsars. In order to reproduce the
complex characteristics  of the red spectra which are 
frequently observed (such as, e.g.,
different power law slopes  in different frequency ranges) most authors
adopt a simple (usually exponential) shot shape, and consider suitable
distributions of shot amplitudes and decay times (see e.g. Belloni and
Hasinger 1990b and  references therein). 
We adopt a somewhat complementary model in which all 
shots have the same shape and amplitude (described by the envelope 
function $g(t)$), such that their  superposition is given by 

\begin{displaymath}
{\tilde f_2}(t) = \sum_j  g(t-t_j) \qquad ,
\end{displaymath}

\noindent
where $t_j$ is the (random) start time of the $j^{th}$ shot. 
This model is more easily handled analytically and yet yields 
model power spectra of arbitrary complexity.

If the aperiodic component is coupled to the periodic one, then
$f_2$ is, at least in part, modulated by $f_1$. Therefore we write 

\begin{displaymath}
f_2(t)= {\bigg [} 1 + \frac{C'}{A} \sum_{n=1}^N B_n 
\sin{(n\omega_0 t + \phi_n)} {\bigg ]} \, {\tilde f_2}(t) \qquad ,
\end{displaymath}

\noindent 
with $C'$ a constant that controls the extent to which the 
periodic signal modulates the shots (the lower $C'$, the lower the
modulation). In order to leave the fluence of each shot unaffected
(at least on average), we have divided the amplitude of the deterministic
function by its mean value $A$. 
Note that $A$ cannot be derived directly from the mean value of the signal,
since this comprises an additional term given by the mean of the shots.
Therefore we define a new constant, $C=C'/A$, which controls the degree
of modulation but is inversely proportional 
to the mean signal. The 
$CB_i$ products provide a measurement of the depth of the shot modulation
in each harmonics.

The Fourier transform of the X-ray pulsar signal 
$f_\infty (t)=f_1(t) + f_2(t) $ is therefore  

\begin{eqnarray}
F_\infty(\omega) &=& A\delta(\omega) + \sum_{n=1}^N {{B_n}\over{2i}}e^
{-i\omega\phi_n}{\big [}\delta(\omega-n\omega_0)-\delta(\omega+n\omega_0)
{\big ]} + \nonumber \\
&\ &+ {\bigg \{}\delta(\omega)+{C\over{2i}}\sum_{n=1}^NB_ne^{-i\omega\phi_n}
{\big [}\delta(\omega-n\omega_0)-\delta(\omega+n\omega_0){\big ]}{\bigg \}}
\ast
{\bigg \{} G(\omega)\sum_je^{-i\omega t_j}
{\bigg \}} \qquad ,\nonumber 
\end{eqnarray}

\noindent 
where $G(\omega)$ is the Fourier transform of the shot envelope function $g(t)$.

For practical applications, the power spectrum must be calculated by
taking into account the effects due to the binning, sampling and finite
duration of the light curves. Following van der Klis (1989) we describe a
finite,  equispaced and binned light curve as (``*" indicates  a convolution) 

\begin{equation}
f(t) = {\big \{ }{\big [} f_\infty (t)w(t) {\big ]}\ast b(t) 
{\big \} } s(t) \label{uno} \qquad ,
\end{equation}

\noindent
where $b(t)$ is the binning function, $s(t)$ the sampling function (a
monodimensional lattice) and $w(t)$ the window function.
By using the convolution theorem, the Fourier transform $F(\omega)$ of 
$f(t)$ writes

\begin{equation}
F(\omega) = 2\pi {\big \{} {\big [} F_\infty (\omega) \ast W(\omega)
{\big ]} B(\omega) {\big \}} \ast S(\omega) \label{due} \qquad ,
\end{equation}

\noindent
where the Fourier transforms of $w(t)$ and $b(t)$ are, respectively,

\begin{eqnarray}
W(\omega) &=& {T\over {2\pi}}{{\hbox{sinc}{\bigg (}{T\over 2}\omega{\bigg )}}}
\sim \delta(\omega) \nonumber \\
B(\omega) &=& {\Delta t\over {2\pi}}{{\hbox{sinc}{\bigg (}{\Delta t
\over 2}\omega {\bigg )}}} \qquad , \nonumber 
\end{eqnarray}

\noindent
where $\sinc (x) = \frac{\sin(x)}x$.
$S(\omega)$ is the reciprocal lattice. $T$ and $\Delta t$ indicate, 
respectively, the light curve duration and its binning time. 
Note that in Eq.~\ref{due} we have used the normalisation 

\begin{displaymath}
F(\omega) = {1\over{2\pi}} \int_{\hbox{\bf R}} f(t) \, e^{-i\omega t}\;
dt \qquad .
\end{displaymath}

The convolution with
$S(\omega)$ in Eq.~\ref{due} can be neglected, as the aliasing
introduced by the sampling is depressed by the $\sinc$ due to the binning.
Moreover the observed power spectra (see Section~\ref{applicazione}) are 
often dominated by the
counting  statistics noise well before the Nyquist frequency, such that 
the low signal to noise further reduces the possibility of detecting any 
alias. The expression for $F_\infty(\omega)$ in Eq.~\ref{due} 
can be simplified by
neglecting  the $\delta$-functions centered on negative frequencies. 
Therefore for $F(\omega)$ we obtain 

\begin{eqnarray}
F(\omega) &=& {\Delta t\over{2\pi}}\hbox{sinc}{\bigg (}{\Delta t\over2}\omega
{\bigg)}
{\Bigg \{}AT\hbox{sinc}
{\bigg (}{T\over2}\omega{\bigg )} +{T\over{2 i}} \sum_{n=1}^N B_n
e^{-in\omega_0\phi_n}\hbox{sinc}{\bigg [}{T\over2}(\omega-n\omega_0)
{\bigg ]} + \nonumber \\
&\ &+ 2 \pi {\bigg \{} G(\omega) \sum_j e^{-i\omega t_j}
 + {C\over{2i}}\sum_{n=1}^N {\bigg [} B_n
e^{-in\omega_0\phi_n} G(\omega -n\omega_0) \sum_j e^{-i(\omega-n\omega_0)t_j}
{\bigg ]}{\bigg \}}{\Bigg \}} \quad , \label{tre}
\end{eqnarray}

\noindent
where $W(\omega)$ was approximated with a $\delta$-function 
in the convolution with $G$, since the shot decay time 
is supposed to be much shorter than the observation duration $T$
(such that $G(\omega)$ is much wider than $W(\omega)$).

The final step is to calculate the power spectrum $P(\omega)$. 
As usual, the simple definition $ P(\omega) = | F(\omega) |^2 $
cannot be adopted as it would include a strong dependence on the 
shot start times. 
Instead the definition 

\begin{equation}
P(\omega) = <| F(\omega) |^2> \label{qua} \qquad 
\end{equation}

\noindent
must be used, where the brackets indicate the average over the ensemble of
realisations of signals with the same deterministic periodic component and 
shot parameters, but with
different $\{t_j\}$. The derivation of $P(\omega)$ is
quite complex, and is described in the Appendix. The result is 

\begin{equation}
\begin{tabular}{lrcl}
(a) \qquad &$P(\omega) $&$\simeq$&$ {{\Delta t^2}\over{4\pi^2}}
\hbox{sinc}^2{\bigg(}{\Delta t\over2}\omega {\bigg)}   {\Bigg\{}
T^2(A+{\bar E})^2\hbox{sinc}^2{\bigg(}{T\over2}\omega {\bigg)} +$ \\
(b) & \ &\ &$+ {T^2\over4}\sum_{n=1}^N B^2_n\hbox{sinc}^2{\bigg [}{T\over2}
(\omega-n\omega_0){\bigg ]} + $\\
(c) & \ &\ &$+ 4 \pi^2 T \nu {\Bigg[} |G(\omega)|^2 + {{C^2}\over{4}}
\sum_{n=1}^N B^2_n |G(\omega-n\omega_0)|^2 {\Bigg]} + $
\\
(d) & \ &\ &$+ C^2 \pi^2 (T^2\nu^2-T\nu) \sum_{n=1}^N
B^2_n |G(\omega-n\omega_0)|^2 \hbox{sinc}^2 {\Big [} {T \over 2}(\omega-n\omega_0)T{\Big]}
 + $\\
(e) & \ & \ &$+ 2  \pi^2 T^2 \nu C  \sum_{n=1}^N
B^2_n |G(\omega-n\omega_0)|^2 \hbox{sinc}^2 {\Big[} {T \over 2}(\omega-n\omega_0)T{\Big]}
{\Bigg\}} \qquad ,$ 
\end{tabular}
\label{cin} 
\end{equation}

\noindent
where $\nu$ is the shot rate, and ${\bar E}$ the mean power in the shot 
component.
It is worth emphasising that the formula above is derived under the 
assumption that the observation duration $T$ is long enough that 
the cross-terms resulting from the relative
phases in the harmonics of the periodic signal can be neglected 
(see the Appendix). The validity of this assumption is verified 
{\it a posteriori} in the application to the observed power spectra 
of selected X-ray pulsars presented in Section~\ref{applicazione}. 

Besides the multiplicative term due to the binning, the right 
hand side of Eq.~\ref{cin} comprises four additive terms; 
these are, respectively: 

(a) The term arising from the average level of the total signal, i.e. the
deterministic periodic signal plus the shot noise. 
This term is sharply peaked around zero frequency. 

(b) The term containing the narrow peaks (width of $\sim 1/T$)
due to the $N$ harmonics of the deterministic periodic signal.

(c) A term consisting of the sum of the red noise spectrum due to the shots' 
envelope ($|G(\omega)|^2$) plus broad wings (width of $\sim \frac1\tau$, 
where $\tau$ is the characteristic decay--time of the shots)
centered around each narrow peak. These 
broad wings are characterised by the same functional form of the red noise
spectrum on both sides of each peak. They arise from the convolution of 
the deterministic signal peaks with the red noise component due to the shots.

(d) An additional term contributing to the narrow peaks 
(width of $\sim 1/T$) centered around the
harmonics. This term is due to the fact that the modulation of all shots 
is in phase with respect to
the deterministic periodic signal. In principle this contribution to 
the narrow peaks could be distinguished from that due to  
the deterministic narrow peaks (term $b$) by virtue of the 
dependence on the shot rate $\nu$. 

(e) A term, contributing to the narrow peaks' amplitude,
which arises from the cross product in the calculation of the squared modulus
of the two terms in Eq.~\ref{tre}. Its amplitude is a factor of $\sim C\nu$
lower than term $d$.

Within our model the presence of broad wings around the peaks (see term $c$)
unambiguously indicates that the shots are modulated by the periodic signal, 
such that a coupling is present. 
Note that the functional form of the broad wings
is dictated by the red noise component, such that the
modelling of the wings (that usually have a poor signal to noise ratio) is 
coupled to the modelling of the red noise component. It 
should be emphasized that no assumption is yet made about envelope function 
of the shots.

The possibility of revealing a modulation of the shots through the 
broad wings of the narrow peaks depends crucially on the ratio of the pulsar 
period to the characteristic decay time $\tau$ of the shots.
To illustrate this point we show in Fig.~\ref{broadpeaks} the model power 
spectra obtained for shots with exponential envelope function 
and decay--time $\tau = 300$~s.
The different curves correspond to 
50\% modulated shots with a sinusoidal signal of period
$P=236,\ 471,\ 942,\ 1885,\ 3770$~s, respectively. 
It is apparent that when $2\pi P \gsim \tau $ the wings are
so broad that they become virtually indistinguishable from
the red noise. This is due to the fact that the shots are too 
short-lived to display a conspicuous periodic modulation.

\section{Application to the Power Spectra of selected X-ray
Pulsars}
\label{applicazione}

\subsection{Fitting procedure}

In order to test the coupling  between the periodic and aperiodic
variability, we  used the model power spectrum in Eq.~\ref{cin} 
to fit the power 
spectra obtained from the X-ray light curves of a few selected X-ray
pulsars. The relevant formula was inserted in the QDP/PLT fitting
program (Tennant 1981), which provides a nonlinear least square fitting
based on the Marquandt method. 
A constant term was added to account for 
the presence of counting statistics (white) noise in the power spectra.  

We used the [4--9] keV light curves of X-ray pulsars obtained with the 
EXOSAT Medium Energy detector array, in consideration of their high
statistical quality and (near) absence of interruptions  
(White and Peacock 1988). The data were extracted from the High Energy
Astrophysics Database System at the Brera Astronomical Observatory (Tagliaferri
\& Stella 1993, 1994, see also White et al. 1995a). For each source a power spectrum was calculated 
over a number of separate observation intervals of the same 
length. For each interval we checked that: (a) the source count rate 
was approximately at the same level; (b) the light curves did not 
contain eclipses, absorption dips or other non-stationary events; 
(c) the position of the narrow peaks due to the periodic pulsar 
signal was the same to within the Fourier resolution 
$1/T$ of the power spectra  (such that the smearing due to the 
orbital Doppler effect and secular period derivative could be neglected).

As customary, we calculated the power spectra of each interval 
after subtracting
the average count rate, such that the contribution of term $a$ in 
Eq.~\ref{cin} can
be neglected and the term itself excluded from the fit. 
These power spectra were then used to
produce an average power spectrum for each X-ray pulsar. The statistical 
error of the power estimates was evaluated based on the standard 
deviation of the average power for each Fourier frequency.  
Note that the interval duration was determined by compromising 
between two conflicting requirements: that the intervals are long 
enough to investigate the low frequency end of the red-noise and  
that their number is high enough (we averaged a number of 7 to 13 
individual spectra) 
to make the distribution 
of the power estimates in the average power spectrum close 
to a Gaussian distribution (as required to apply standard least 
square fitting techniques, see e.g. Israel \& Stella 1996 and references 
therein).   

In the EXOSAT database we identified three X-ray pulsars, with
observations meeting the above requirements. These are 
Vela~X-1 ($P=282.6$~s), 
4U1145-62 ($P=292.1$~s) and Cen~X-3 ($P=4.8$~s) (see e.g. Nagase 1989;
White et al. 1995b and references therein). Details of the 
relevant observations are given in Table~\ref{obstab}, together with the 
interval length and binning time adopted in our analysis. 

The fitting procedure for the average power spectrum of each X-ray pulsar 
consists of three steps.  

(i)  First we fit our model under the hypothesis that no coupling is 
present between the deterministic periodic signal and the shots. This 
is obtained by setting $C=0$. Consequently terms $d$ and $e$  and the 
2nd part of term $c$ of Eq.~\ref{cin} 
are excluded from the fit. Therefore this
case  includes only the red noise component and
the ``narrow peaks" (under the assumption that the 
effects of the relative phases between the harmonics can be neglected).
For $|G(\omega)|^2$ we find that a King-like model of 
the form  

\begin{equation}
|G(\omega)|^2 = D {\bigg [} 1+{\Big (}{\omega\over\omega_c}{\Big )}^2
{\bigg]}^{-\alpha} \label{redn} \qquad 
\end{equation}

\noindent
reproduced quite accurately the power law-like  behaviour and the low 
frequency flattening that characterizes the red noise spectra we analysed.

(ii) The model used in step (i) is generalised to include the 
effects due to the relative phases between the harmonics. The relevant
formula is given by Eq.~11 in Angelini et al. (1989). This step 
is designed to provide an {\it a posteriori}
check of the assumptions under which Eq.~\ref{cin} has been derived. 

(iii) In the third step the model that includes the coupling 
between the periodic signal and the red noise is used. In this case the coupling
constant $C$ of Eq.~\ref{cin} is used as a free parameter in the fit.  
The only 
contribution to the broad peak wings derives from the second part of term $c$
in Eq.~\ref{cin}. Note that the shape of the broad wings depends on   
$|G(\omega)|^2$, which in turn is mainly determined by the shape of the 
red noise. In practical applications of Eq.~\ref{cin},    
it is found that the signal to noise of the observed power spectra is
insufficient to isolate the contribution to the
narrow peaks deriving also from terms $d$ and $e$ of Eq.~\ref{cin}. 
Therefore we 
carry out the fit by neglecting these two terms. In this approach the shot
rate cannot be obtained directly from the fit to the power spectrum. Moreover
the value of $C$, which is related to the ratio between the height of 
the narrow peak and broad wing component, is underestimated. This is because,
by neglecting terms $d$ and $e$ in Eq.~\ref{cin},
the height of the narrow peaks is attributed only to the deterministic component.
A value of $C$ significantly different from zero would nevertheless 
reveal the presence of a coupling.  

\subsection{Vela X-1}

The average power spectrum of Vela X-1 obtained from the 4-9~keV EXOSAT 
light curves is shown in Fig.~\ref{velaspec}. The fit obtained under the 
hypothesis
of no coupling (step $i$) is shown in the upper panel of Fig.~\ref{velaspec}, 
while 
the corresponding residuals are shown in the lower panel. The fit 
to the red noise provides a King model 
index of $\alpha \simeq 0.58$, whereas the break frequency
$\nu_c=\omega_c/2\pi$ was too low to be measured. The fit included the first 
eleven harmonics of the pulse frequency 
$\nu_p=(3.5379 \pm 0.0007)\times 10^{-3}$~Hz.  The corresponding $\chi^2$
is 325.9 for 189 degrees of freedom (hereafter {\it dof}). 
It is apparent that the peak wings are not well modelled.  
Including the effects of the relative phases between the harmonics 
(step $ii$) left the best fit and corresponding $\chi^2$ (=323.3)
virtually unaffected, while the number of {\it dof} decreased to 180. An F-test
for the  addition of 9 free parameters, confirmed that the improvement of the
fit is not significant. 

Having checked that the relative phases between the harmonics 
can be neglected, we proceeded to step $iii$ by 
allowing for a periodic modulation of the
shots (the coupling constant $C$ is now treated as a free parameter). 
A much better fit is obtained in this case with
$\chi^2/dof = 229.3/188$. This corresponds to an F-test chance probability 
of $<10^{-15}$ for the addition of one free parameter with respect to the 
fit of step $i$. Fig.~\ref{velaspec}(b) shows that the new fit reproduces far more 
accurately the broad wings of the peaks. 
For the red noise component, the new fit  provides a break frequency of 
$\nu_c = (2.0 \pm 1.3) \times 10^{-4} Hz << \nu_p$ and a King model 
index of $\alpha = 0.72 \pm 0.05$ (errors are $90 \%$ confidence).
The products $CB_n$ of the coupling constant and the
harmonics' amplitude obtained from the fit are given in Table~\ref{cbntab}.

It is useful to define a parameter $R = C \sum_{n=1}^N B_n$
which is related to the degree of modulation of the pulse. 
In the case of a sinusoidal signal, $R$ represents
the relative depth of the periodic modulation of the shots. 
For a periodic signal containing more than one harmonics, the  
shape of the shot modulation cannot be reconstructed due to the 
absence of information about the relative phases of the harmonics. 
Consequentely in this case $R$ does not measure the relative modulation 
of the shots and values of $R > 1$ can be obtained. Nevertheless the
value of $R$  provides an indication of whether the shots are weakly 
($R \ll 1$) or strongly ($R\approx 1$) modulated.  
For Vela X-1 we find a large value of $R = 3.2 \pm 0.2$, 
pointing to a  large amplitude of the shot modulation.
   
\subsection{4U1145-62}

The average power spectrum of 4U1145-62 was obtained from the X-ray 
light curves of separate
EXOSAT observations, since no single long observation was available
(cf. Table~\ref{obstab}). The fit obtained under the hypothesis
of no coupling between the periodic and the red noise (step $i$) is shown in
the upper panel of Fig.~\ref{4uspec}(a). 
The first five harmonics of the pulsar modulation 
at $\nu_p=(3.4229 \pm 0.0006)\times 10^{-3}$~Hz were included in the fit. 
Note that in this case the red-noise component
extends in a power law fashion down to the lowest frequencies sampled
in the power spectrum ($\alpha = 0.41\pm 0.02$),
such that only an upper limit to the break frequency
$\nu_c \lsim 10^{-4}$~Hz could be obtained. This frequency is  $\ll \nu_p$, 
like in the case of Vela~X-1.  For step $i$ and 
step $ii$ fits a $\chi^2$ of 299 was obtained, respectively for 
189 and 180 $dof$. Therefore also in this case the effects of the 
relative phases between the harmonics are negligible. 
When the coupling constant $C$ was allowed to vary (step $iii$) we 
find a $\chi^2/dof$ of 275/188; correspondigly the F-test chance probability 
for the addition of one free parameter relative to the step $i$ fit was 
found to be $\sim 3\times 10^{-5}$. Therefore we conclude that also in the 
case of 4U1145-62 the shots are modulated. The values of $CB_n$ are given 
in Table~\ref{cbntab}, while for $R$ we derive a value of $1.6\pm0.2$.  

\subsection{Cen X-3}

The power spectrum of the EXOSAT light curves of 
Cen~X-3 is different from the two cases 
presented above in that the red noise  component shows a clear flattening for
frequencies shorter than the fundamental of the periodic modulation (see
Fig.~\ref{censpec}), implying that the shot duration is comparable to the 
X-ray pulsar period. Under these circumstances, any modulation 
of the shots with the periodic signal would give rise to such broad wings
around the power spectrum peaks, that they would be hardly distiguishable 
from the red noise itself (see Section~\ref{modello}). 
We confined our analysis of the power spectrum of Cen~X-3 to the
range of frequencies  between 0.02 and 0.8~Hz, such that the King-like model of
Eq.~\ref{redn} provided a reasonably accurate fit of the red-noise component. 
The first three harmonics of the periodic modulation at 
$\nu_p = (2.0692 \pm 0.0008) \times 10^{-1}$~Hz were included in the fit. 

Under the hypothesis of no coupling, the fit (step $i$) gives a 
$\chi^2/dof$ of 2108/1589, for $\alpha= 0.43 \pm 0.03$
and $\nu_c = 1.2 \pm 0.1$~Hz. 
The step $ii$ fit produced the same $\chi^2$, for 1580 $dof$; the effects of 
the relative phases between harmonics are therefore negligible also 
in the case of Cen~X-3. 
Allowing for a periodic modulation of the shots (step $iii$) yielded a fit  
with a $\chi^2$ of 2108, for 1588 $dof$. This shows that no broad 
wings are detected around the peaks in the case of Cen~X-3
(see Table~\ref{cbntab} for upper limits on the $CB_n$). Note however that
since the expected  shape of the wings is dictated by the shape of the
red-noise (see above), only a poor upper limit is found for $R$  ($ \lsim 0.66$ 
at the
90\% confidence  level). This indicates that a fairly strong coupling 
between the periodic and aperiodic variability might 
be present and remain undetected in Cen~X-3.

\section{Discussion}
\label{discussione}

Based on a simple model consisting of the sum of a periodic signal 
plus random shots characterised by an arbitrary degree of modulation with
the  periodic signal, we have shown that in two out of
three X-ray  pulsars that we have analysed (Vela X-1 and 4U1145-62)
there exists a strong  
coupling between the periodic and red noise variability. This coupling 
is revealed through the broad wings that are found around 
the power spectrum peaks arising from the periodic X-ray pulsar 
signal. As the shape of the red noise 
component dictates the shape of the broad wings, these are more easily
detected when the red noise power increases shortwards of the 
pulsar frequency $\nu_p$.
This is indeed the case for Vela X-1 and 4U1145-62. On the 
contrary, the red noise component of Cen~X-3  flattens 
around $\nu_p$, such that any wings would be so broad 
that they are very difficult to separate from the red-noise component. 
We find no evidence for such very broad wings around the power spectrum peaks 
of Cen~X-3; the corresponding upper limit to the modulation of 
the shots is poor and still allows for a relatively strong coupling.
This is likely the case also in other disk-fed X-ray pulsars that
show unambiguously a flat-topped red noise component steepening 
above frequencies of $\gsim \nu_p$. 

Our results therefore suggest that a coupling between the periodic 
and red noise variability might be frequently present in X-ray pulsars. 
This is not a surprising result as one would expect that any 
accretion flow inhomogeneity (or ``shot") responsible for the red noise,
produces  most of its X-ray luminosity in the accretion column close to 
the neutron star surface. The combination of rotation and 
radiative transfer effects should therefore produce a periodic modulation
of the shots similar to that of any continuum X-ray accretion 
onto the polar caps. 
If, as assumed in our model, the modulation across different shots is phase
coherent, then narrow power spectrum peaks are also generated by the 
shot component (see Section~\ref{modello}). One might speculate that the whole
X-ray flux is produced by the superposition of random modulated shots ({\it
i.e.} no deterministic  periodic signal is present). 
To ascertain whether this is the case
is beyond the scope of the present paper.

There is also an interesting consequence of our work 
concerning the  correlation between the pulsar frequency and the 
knee frequency in the power spectrum continuum 
that was reported by Takeshima (1992). 
It is apparent from the power spectrum of Vela~X-1 in Fig.~\ref{velaspec}
that the
broad wings around the peaks might mimic the presence of a break
in the red noise  component around $\nu_p$. To address this point in a 
quantitative fashion we adopted the power spectrum model used by Takeshima 
(1992). This comprises the sum of: (a) the narrow peaks from the 
periodic component; (b) a power law; (c) a flat component followed by a 
power law above a knee frequency $\nu_{knee}$. The fit to the 
power spectrum of Vela~X-1 obtained in this way is given in the upper panel
of Fig.~\ref{giapspec}, showing also the 
separate contributions from the three model components.
A value of $\chi^2/dof$ of $297/186$ was obtained; this is considerably worse 
than the fit based on our modulated shot noise model ($\chi^2/dof = 229/188$).
The bottom panel of Fig.~\ref{giapspec} shows also the latter fit, with the 
separate contributions from the red noise, the narrow peaks and the broad 
wings. 
It is apparent that the broad wings provide a much more accurate fit 
of the power spectrum in between the narrow peaks. We conclude that the 
red noise power spectrum of Vela~X-1 shows no evidence of a knee around 
$\nu_p$. A similar conclusion, though with a lower statistical significance,
is reached also for the power spectrum of 4U~1145-62. In this
case we obtained a $\chi^2/dof$ of 299/189, 284/186 and 272/188 for 
the step (i) fit, the Takeshima model and the step (iii) fit, respectively.

By analogy, this suggests that also for other X-ray
pulsars with red noise components increasing shortwards 
of $\nu_p$, the broad wings resulting from the modulated shots might mimick
the additional component with a knee at $\nu_{knee} \sim \nu_p$ 
that was introduced by Takeshima (1992). 
Therefore this component is probably not required 
and a red noise knee, if present, would be at 
$\nu_{knee} \ll \nu_p$. On the other hand, the case 
of Cen~X-3 is probably typical of those X-ray pulsars for which the 
red noise flattens indeed below frequencies of $\nu_{knee} \gsim \nu_p$;
any broad wings would be (nearly) undetectable and their contaminating 
effect on the power spectrum continuum negligible. 
The discussion above indicates that, if the power spectrum 
model does not include the broad wings around the periodic modulations peaks, 
a bias towards the $\nu_{knee} \sim \nu_p$ relation is likely introduced.

This is but an example of the caution that should be used in isolating 
the continuum power spectrum components that arise {\it only} from the 
aperiodic variability of X-ray pulsars. Indeed the peaks' broad wings 
originating from the coupling of the periodic and aperiodic variability, 
if not adequately modelled, can lead to inaccurate conclusions concerning 
the shape, amplitude and frequency range of the continuum power spectrum
components. 
A systematic reanalysis of the Ginga power spectra of X-ray pulsars should be
carried  out in the light of our present work. Valuable new information 
should derive from the X-ray pulsar power spectra that are being
obtained by the RXTE. 
\acknowledgements
LS acknowledges useful discussions with L. Angelini. This work was partially 
supported through ASI grants. 

\newpage
\appendix
\section{Appendix}
\label{appendice}

The derivation of Eq.~\ref{cin} begins with
the application of definition~\ref{qua} to Eq.~\ref{tre}.
By averaging over the ensemble of different realisations of the shots start
times $\{t_j\}$, we obtain:
\begin{eqnarray}
P(\omega) &=& {\Delta t^2 \over{4\pi^2}}\hbox{sinc}^2{\bigg (}{\Delta t
\over2}\omega {\bigg)}{\Big <}{\Bigg |}
AT\hbox{sinc}
{\bigg (}{T\over2}\omega{\bigg )} +{T\over{2 i}} \sum_{n=1}^N B_n
e^{-in\omega_0\phi_n}\hbox{sinc}{\bigg [}{T\over2}(\omega-n\omega_0)
{\bigg ]} + \nonumber \\
&\ &+ 2\pi{\bigg \{}   G(\omega) \sum_j e^{-i\omega t_j}
 + {C\over{2i}}\sum_{n=1}^N {\bigg [} B_n 
e^{-in\omega_0\phi_n} G(\omega -n\omega_0) \sum_j e^{-i(\omega-n\omega_0)t_j}
{\bigg ]}{\bigg \}}{\Bigg |}^2{\Big >} \qquad .
\nonumber
\end{eqnarray}
As $T\omega_0$ is $\gg1$, the interference terms between different
harmonics can be neglected. In this limit the $\sinc$ function is approximated
by a $\delta$-function. A consequence of this is that 
also the relative phases between the harmonics are neglected. As described in 
Section~\ref{applicazione}, this assumption is verified {\it a posteriori}
in the fitting procedure. We therefore obtain:

\begin{equation}
\begin{tabular}{ll}
(a)& $P(\omega) = {\Delta t^2 \over{4\pi^2}}\hbox{sinc}^2{\bigg (}
{\Delta t\over2}\omega
{\bigg)}
{\Bigg \{}A^2T^2\hbox{sinc}^2
{\bigg (}{T\over2}\omega{\bigg )} +{T^2\over 4} \sum_{n=1}^N B^2_n
\hbox{sinc}^2{\bigg [}{T\over2}(\omega-n\omega_0)
{\bigg ]} +$ \\
(b) & $\quad + 4\pi^2 {\Big <}{\bigg |} G (\omega) \sum_j e^{-i\omega t_j}
 + {C\over{2i}}\sum_{n=1}^N {\bigg [} B_n 
e^{-in\omega_0\phi_n} G(\omega -n\omega_0)\sum_j e^{-i(\omega-n\omega_0)t_j}
{\bigg ]}{\bigg |}^2{\Big >} +$ \\
(c) & $\quad + 4\pi AT \hbox{sinc}{\bigg (}{T\over2}\omega{\bigg )} {\bf{Re}}
{\bigg[} G(\omega) <\sum_j e^{-i\omega t_j}>{\bigg]} + 
$\\
(d) & $\quad + \pi TC \sum_{n=1}^N B^2_n\cos(n\omega_0\phi_n)
\hbox{sinc}{\bigg [}{T\over2}(\omega-n\omega_0){\bigg]}
{\bf{Re}}{\bigg[}e^{-in\omega_0\phi_n} G(\omega-n\omega_0)
<\sum_j e^{-i(\omega-n\omega_0)t_j}>{\bigg ]}+$\\
(e) & $\quad + \pi TC \sum_{n=1}^N B^2_n\sin(n\omega_0\phi_n)
\hbox{sinc}{\bigg [}{T\over2}(\omega-n\omega_0){\bigg]}
{\bf{Im}}{\bigg[}e^{-in\omega_0\phi_n} G(\omega -n\omega_0) 
<\sum_j e^{-i(\omega-n\omega_0)t_j}>{\bigg ]} {\Bigg \}} \qquad . $
\end{tabular}
\label{sei}
\end{equation}
Note that the cross products in term $b$ cannot be neglected because of the 
considerable width of $G(\omega)$. However it can be shown that 
these cross products cancel each other exactly over the ensemble average.
Term~\ref{sei}$b$ becomes
\begin{displaymath}
4\pi^2 {\Bigg\{} |G(\omega)|^2 <{\big |}\sum_j e^{-i\omega t_j}{\big |}^2>
 + {C^2\over{4}}\sum_{n=1}^N {\bigg [} B^2_n |G(\omega-n\omega_0)|^2
<{\big|}\sum_j e^{-i(\omega-n\omega_0)t_j}{\big|}^2>
{\bigg ]}{\Bigg \}} \qquad .
\end{displaymath}
In order to calculate $<{\big |}\sum_j e^{-i\alpha\theta_j}
{\big |}^2>$, we separate real and imaginary parts of the sum
and work out the ensemble average separately.
We obtain
\begin{displaymath}
<{\Big |}\sum_j e^{-i\alpha\theta_j}{\Big |}^2> = N +
(N^2-N)(<\cos\alpha\theta_j>^2 + <\sin\alpha\theta_j>^2) \qquad .
\end{displaymath}
Note that, unlike the case of a random walk, the second terms become very
large close to the resonance condition $\alpha=0$ (i.e. $\omega=n\omega_0$
in Eq.~\ref{sei}).
If the mean over the ensemble is equal
to the mean over time (i.e. the system is ergodic),
we have 
\begin{displaymath}
<f> = {\bar f} = {1\over T}\int_{-{T\over 2}}^{T\over2}f(t)\,dt \qquad ,
\end{displaymath}
\noindent
and  
\begin{displaymath}
<{\Big |}\sum_j e^{-i\alpha\theta_j}{\Big |}^2> = N +
(N^2-N)\, \hbox{sinc}^2{\bigg(}{{\alpha T}\over 2}{\bigg)}  \qquad .
\end{displaymath}
Term~\ref{sei}$b$ therefore becomes:

\begin{eqnarray}
&\ & 4\pi^2 T\nu {\Bigg [} |G(\omega)|^2 + {C^2\over{4}}
\sum_{n=1}^N {B^2_n} |G(\omega-n\omega_0)|^2 {\Bigg]} + \nonumber \\
&\ &\pi^2 C^2 (T^2\nu^2-T\nu)\sum_{n=1}^N
B^2_n\hbox{sinc}^2{\Big(} {{(\omega-n\omega_0)T}\over2}{\Big)}
|G(\omega-n\omega_0)|^2 \qquad . \nonumber
\end{eqnarray}

Term~\ref{sei}$c$ represents the contribution of the shot average to the 
zero frequency power, while terms~\ref{sei}$d$ and~\ref{sei}$e$ 
are simplified in a manner similar to~\ref{sei}$b$. 
We finally obtain Eq.~\ref{cin}.

\newpage

\begin{table}
\caption{{EXOSAT ME Observations }\label{obstab}}
\bigskip
\centerline{
\begin{tabular}{c|ccccccc}
Name      & Time & Seq.     &  Exposure    & Count Rate & $^aN_{int}$ & $^bT_{int}$ & $\Delta t$ \\ 
\         & (yy.ddd) & \    &  (s)         & (cts/s)    &           &   (s)     & (s) \\ \hline \hline
Vela X-1  & 85.044   & 1402 &  37341       & $73.2 $ & 7   & 5120      &  5  \\ \hline
4U1145-62 & 85.001   & 1324 &  34000       & $42.0 $ & 13  & 12400     &  10 \\
          & 85.002   & 1326 &  21220       & $35.1 $ &     &           &     \\
          & 85.004   & 1329 &  24070       & $35.0 $ &     &           &     \\
          & 85.005   & 1332 &  23570       & $34.8 $ &     &           &     \\
          & 85.008   & 1339 &  25660       & $21.1 $ &    &           &     \\ \hline
Cen X-3   & 85.193   & 1697 &  40685       & $315.1$ & 11 & 2048      &  0.5 
\end{tabular}
}
{\footnotesize
$^a$ Number of intervals used to calculate the average power spectrum.

$^b$ Duration of each interval.}

\end{table}

\begin{table}
\caption{{Results from step $iii$ fits. 
}\label{cbntab}}
\bigskip
\centerline{
\begin{tabular}{c|ccc}
\ & Vela X-1 & 4U1145-62 & Cen X-3 \\ \hline \hline 
$R       $& $3.2  \pm 0.2           $ & $1.6 \pm 0.2            $ & $< 0.66$ \\
$CB_1   $ & $0.61^{+0.13}_{-0.11}   $ & $0.79^{+0.17}_{-0.17}   $ & $< 0.37$ \\
$CB_2   $ & $0.63^{+0.10}_{-0.09}   $ & $0.29^{+0.09}_{-0.08}   $ & $< 0.23$ \\
$CB_3   $ & $0.10^{+0.03}_{-0.03}   $ & $0.07^{+0.03}_{-0.02}   $ & $< 0.06$ \\
$CB_4   $ & $0.41^{+0.07}_{-0.06}   $ & $0.25^{+0.07}_{-0.06}   $ & - \\
$CB_5   $ & $0.44^{+0.07}_{-0.06}   $ & $0.17^{+0.05}_{-0.05}   $ & -\\
$CB_6   $ & $0.22^{+0.04}_{-0.03}$ & - & - \\
$CB_7   $ & $0.26^{+0.05}_{-0.04}$ & - & - \\
$CB_8   $ & $0.14^{+0.03}_{-0.03}$ & - & - \\
$CB_9   $ & $0.10^{+0.03}_{-0.03}$ & - & - \\
$CB_{10}$ & $0.11^{+0.03}_{-0.02}$ & - & - \\
$CB_{11}$ & $0.14^{+0.03}_{-0.02}$ & - & -
\end{tabular}
}
\end{table}

\newpage

\begin{figure}
\caption{{Model power spectra (cf. Eq.~(5)) for shots with an exponential 
envelope function of decay time $\tau = 300$~s. The shots are 50\% 
modulated with a sinusoidal signal. Different curves corresponds 
to different values of the period P; from the top to the bottom this is 
236, 471, 942, 1885 and 3770~s. The adopted light 
curve duration and binning time were 
512000~s and 5~s, respectively. 
It is apparent that the width of  broad wings below the sharp peak 
increases as P decreases.} \label{broadpeaks}}

\caption{{Average power spectrum obtained from the EXOSAT
light curves of Vela~X-1, together with the best fit models 
obtained in the case of non-modulated shots (step $i$, panel $a$) 
and modulated shots (step $iii$, panel $b$). The lower part of each 
panel shows the corresponding residuals.}\label{velaspec}}

\caption{{Average power spectrum obtained from the EXOSAT
light curves of 4U~1145-62, together with the best fit models 
obtained in the case of non-modulated shots (step $i$, panel $a$) 
and modulated shots (step $iii$, panel $b$). The lower part of each 
panel shows the corresponding residuals.}\label{4uspec}}

\caption{{Average power spectrum obtained from the EXOSAT
light curves of Cen~X-3, together with the best fit models 
obtained in the case of non-modulated shots (step $i$) 
and modulated shots (step $iii$). In this case the two fits are virtually
identical. The lower part of the figure shows the corresponding 
residuals.}\label{censpec}}

\caption{{Comparison between the best fit to the average spectrum of Vela X-1 
obtained with model of Takeshima (panel $a$) and the modulated shot model
(step $iii$, panel $b$). The contribution from the three components
of each model is also plotted. The lower part of each 
panel shows the corresponding residuals.}\label{giapspec}}

\end{figure}

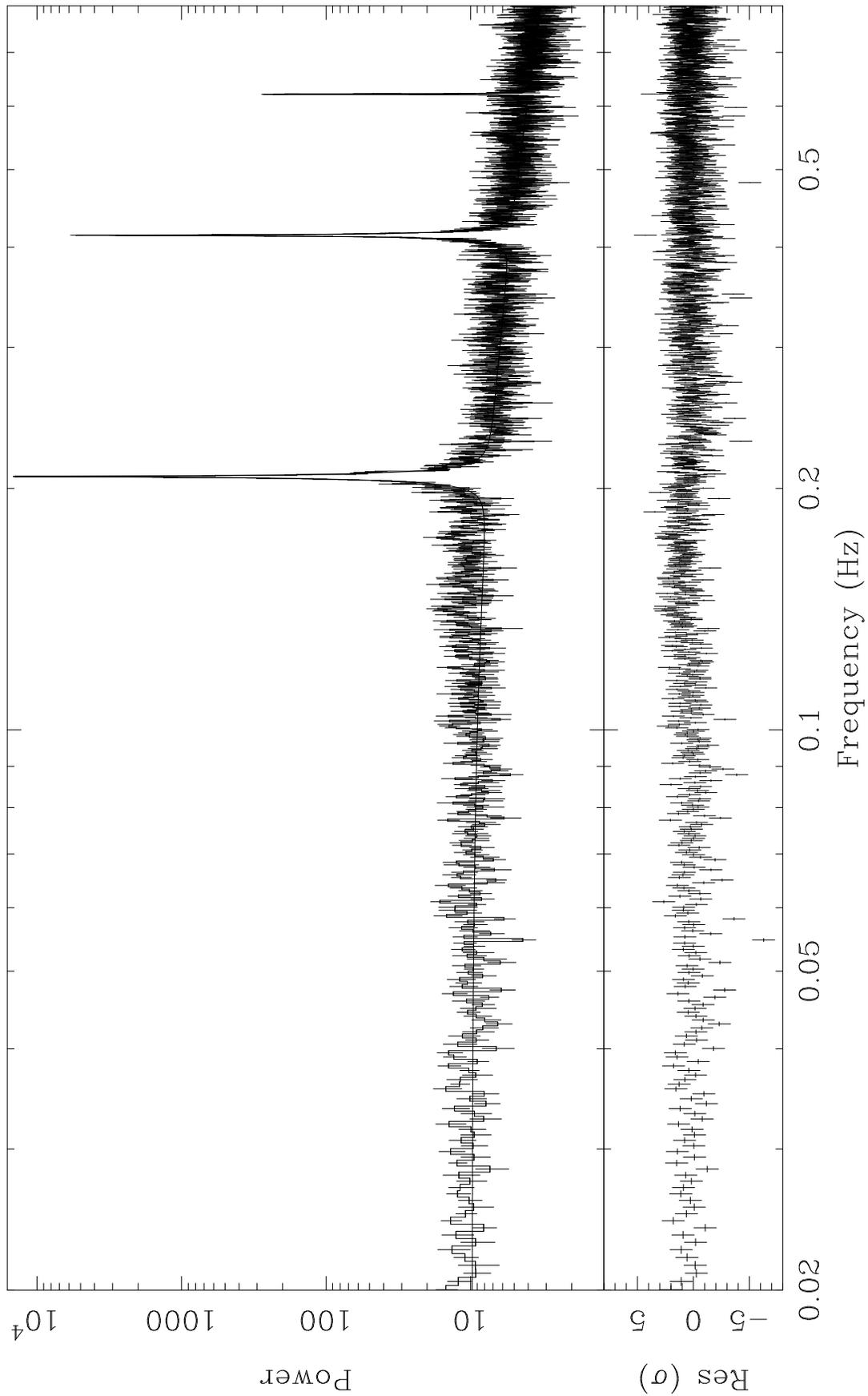
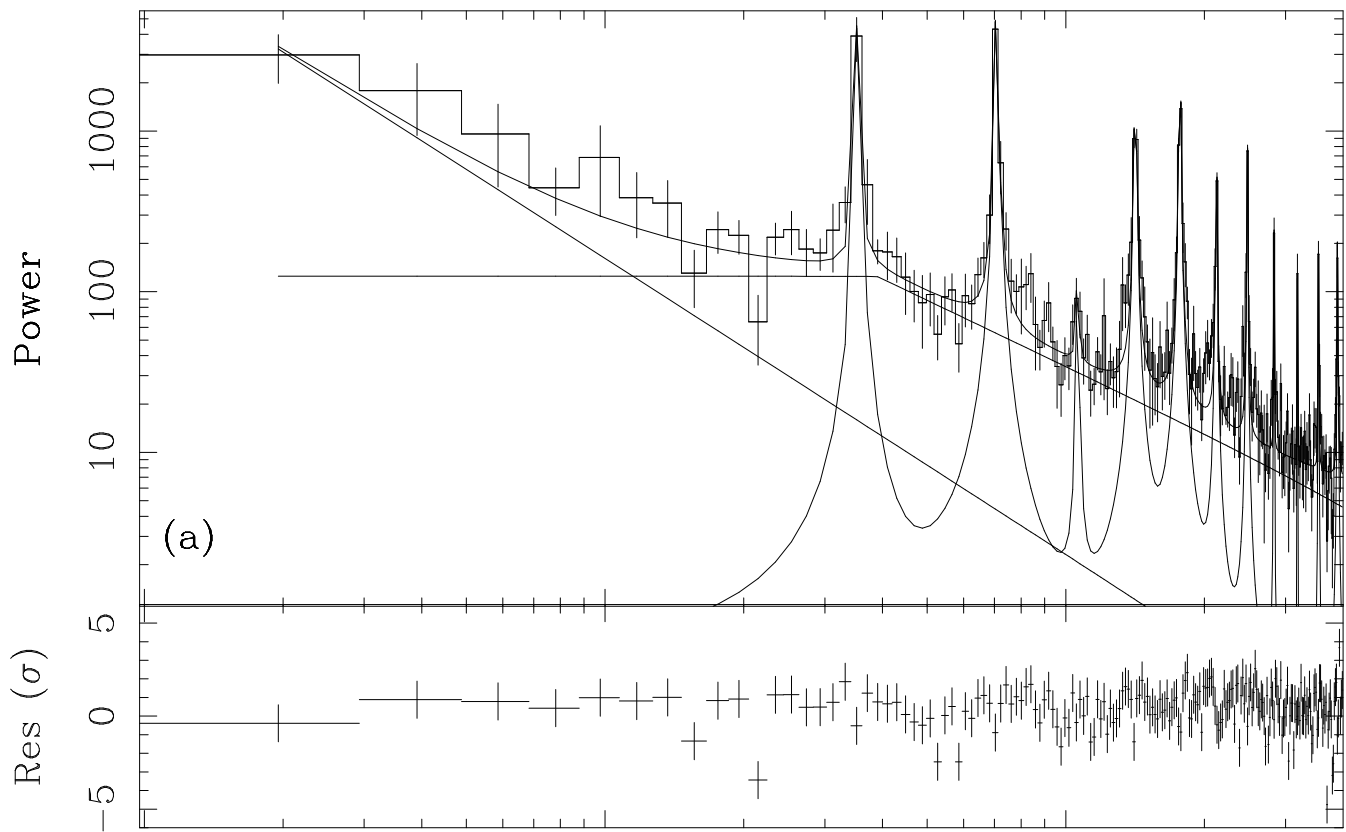
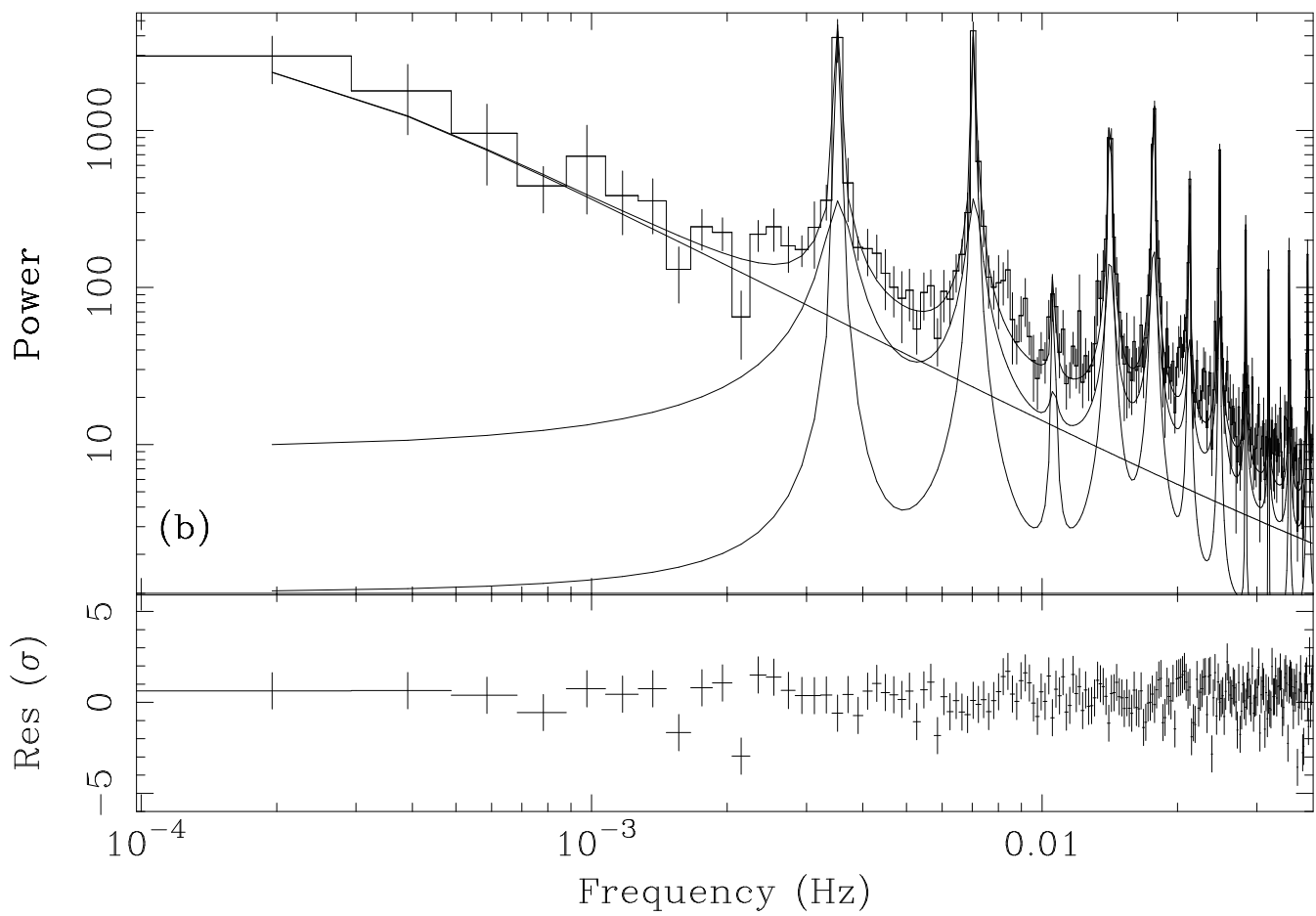
\end{document}